\documentclass[sigconf, nonacm]{acmart}
\usepackage{balance}
\usepackage[shortend,vlined,ruled]{algorithm2e}
\usepackage{amsthm}
\usepackage{siunitx}
\usepackage{soul}
\usepackage{xargs}
\usepackage[conf={restate,no link to proof}]{proof-at-the-end}
\usepackage{enumitem}

\usepackage{tikz} 
\usepackage{pgfplots}
\pgfplotsset{compat=newest} 
\usepgfplotslibrary{units} 
\usetikzlibrary{pgfplots.groupplots}
\usepackage{array}

\newcommand\vldbdoi{XX.XX/XXX.XX}
\newcommand\vldbpages{XXX-XXX}
\newcommand\vldbvolume{14}
\newcommand\vldbissue{1}
\newcommand\vldbyear{2020}
\newcommand\vldbauthors{\authors}
\newcommand\vldbtitle{\shorttitle} 
\newcommand\vldbavailabilityurl{URL_TO_YOUR_ARTIFACTS}
\newcommand\vldbpagestyle{plain}

\def\notationcolor{blue} 

\newcommand{\notation}[2]{\newcommand{#1}{{\textcolor{\notationcolor}{\ensuremath{#2}}}}}

\notation{\mech}{M} 
\notation{\randalg}{\mathcal{A}} 
\notation{\data}{D} 
\notation{\truedata}{\data_{\text{act}}} 
\notation{\att}{\mathbf{A}} 
\notation{\rec}{r} 
\notation{\outp}{\omega}
\notation{\globalsens}{\Lambda}
\notation{\localsens}{\Lambda^s}
\notation{\neigh}{\mathcal{N}}
\notation{\query}{q}
\notation{\npairs}{NPairs}
\notation{\hint}{h}
\notation{\chooser}{Chooser}

\newcommandx*\pquery[2][1=\ensuremath{\phi},2=b]{\textcolor{\notationcolor}{\ensuremath{\query_{#1,#2}}}}

\newcommandx*\tquery[4][1=i,2=b,3=u,4=v]{\pquery[{\att_{#1}\in [#3, #4)}][#2]}

\newcommandx*\mquery[4][1=i,2=b,3=u,4=v]{\textcolor{\notationcolor}{\ensuremath{\mech^{(#1)}_{[#3,#4),#2}}}}

\usepackage{hyperref}
\usepackage{balance}
\begin{document}
\title{Reconstruction Attacks on Aggressive Relaxations of Differential Privacy}


\author{Prottay Protivash}
\affiliation{%
  \institution{Pennsylvania State University}
  \city{State College}
  \state{PA}
}
\email{pxp945@psu.edu}

\author{John Durrell}
\affiliation{%
  \institution{Pennsylvania State University}
  \city{State College}
  \state{PA}
}
\email{jmd6968@psu.edu}

\author{Zeyu Ding}
\affiliation{%
  \institution{Pennsylvania State University}
  \city{State College}
  \state{PA}
}
\email{zyding@psu.edu}

\author{Danfeng Zhang}
\affiliation{%
  \institution{Pennsylvania State University}
  \city{State College}
  \state{PA}
}
\email{zhang@cse.psu.edu}

\author{Daniel Kifer}
\affiliation{%
  \institution{Pennsylvania State University}
  \city{State College}
  \state{PA}
}
\email{dkifer@cse.psu.edu}




\begin{abstract}
Differential privacy is a widely accepted formal privacy definition that allows aggregate information about a dataset to be released while controlling privacy leakage for individuals whose records appear in the data. Due to the unavoidable tension between privacy and utility, there have been many works trying to relax the requirements of differential privacy  to achieve greater utility.

One class of relaxation, which is starting to gain support outside the privacy community is embodied by the definitions of \emph{individual differential privacy} (IDP) and \emph{bootstrap differential privacy} (BDP). The original version of differential privacy defines a set of neighboring database pairs and achieves its privacy guarantees by requiring that each pair of neighbors should be nearly indistinguishable to an attacker. The privacy definitions we study, however, aggressively reduce the set of neighboring pairs that are protected.

Both IDP and BDP define a measure of ``privacy loss'' that satisfies formal privacy properties such as postprocessing invariance and composition, and achieve dramatically better utility than the traditional variants of differential privacy. However, there is a significant downside -- we show that they allow a significant portion of the dataset to be reconstructed using algorithms that have arbitrarily low privacy loss under their privacy accounting rules.

We demonstrate these attacks using the preferred mechanisms of these privacy definitions. In particular, we design a set of queries that, when protected by these mechanisms with high noise settings (i.e., with claims of very low privacy loss),  yield more precise information about the dataset than if they were not protected at all.

\end{abstract}

\maketitle

\section{Introduction}\label{sec:intro}
 Statistical agencies face the challenge of releasing data products that are detailed and statistically useful while also meeting the legal and ethical obligations to protect the confidentiality of individuals providing the data. Similarly, companies seek to gain a competitive advantage by mining detailed information about their user base while still providing confidentiality guarantees to those users.
 
In some areas, differential privacy \cite{dmns06,ourdata,zcdp,renyidp} is gaining acceptance as a source of viable solutions to these problems \cite{rappor,prochlo,applediffp,DingKY17,elasticsensitivity,ashwin08:map,onthemap,Haney:2017:UCF,abowd18kdd}. However, the use of differential privacy to protect Census data has also drawn fierce criticism, most recently with a group of prominent economists and statisticians calling for the Census Bureau to stop using it \cite{hotzetal}.
Such reactions are often due to frustration with the tension between utility and privacy. 
For example, differential privacy has many known mathematical lower bounds that clearly delineate the accuracy with which information can be released at a given level of privacy (see, for example, \cite{Balcer_Vadhan_2019,Vadhan2017,geomdp,twopartydp,DN03,shivalearn,abowd2021an,SteinkeU17}).
 
One the one hand, similar restrictions (hence similar criticisms) would apply to \emph{any} method that protects confidentiality -- producing ``privacy-protected'' data products that allow arbitrary analyses to be conducted accurately will result in reconstruction of nearly all of the underlying confidential data \cite{DN03}, and hence would provide no confidentiality.
However, the trade-off between privacy and utility in practical applications is still very much an open question, and this has led to many relaxations of differential privacy (see \cite{sokdps} for an exhaustive survey). 

In this paper, we study (and develop attacks for) a type of privacy definition that re-examines the concept of \emph{neighboring databases} that is fundamental to differential privacy. Informally, differential privacy seeks to ensure that a data release mechanism $\mech$ behaves ``similarly'' on databases $\data_1$ and $\data_2$ when they are ``neighbors'' of each other. Intuitively, this means that $\mech$ masks the differences between $\data_1$ and $\data_2$. Thus, if neighboring datasets are defined to be all pairs of datasets that differ on the value of a record, this definition provides plausible deniability: an attacker would not be able to determine the contents of any target individual's record since the behavior of $\mech$ would be almost unrelated to the actual contents of the record.
 
 Relaxations of differential privacy that target the definition of neighbors seek to change what a mechanism $\mech$ attempts to hide. The particular class of relaxations \cite{idp,Bootstrap} we are interested in, which we call \emph{empirical neighbors}, argue that if $\data_1$ and $\data_2$ are unrelated to the actual dataset $\truedata$ that will be the input $\mech$, why should $\mech$ be designed to hide the differences between $\data_1$ and $\data_2$ \cite{idp,Bootstrap}?
 Instead, such proposed relaxations try only to hide the differences between the actual dataset  and some suitable alternatives.
 
For example, \emph{Individual Differential Privacy} (IDP) \cite{idp} (\emph{not} to be confused with personalized differential privacy \cite{pdp1,pdp2}) considers two databases $\data_1,\data_2$ to be neighbors if one of them is the actual dataset $\truedata$ owned by a statistical agency and the other can be obtained from $\truedata$ by modifying a single record.
Their argument is that this is precisely what statistical agencies need because it provides plausible deniability of any record in $\truedata$ (and hence any additional protections provided by differential privacy are unnecessary). 
This rationale sounds convincing to many outside the privacy community. For example Hotz et al. \cite{hotzetal} called for a moratorium on the use of differential privacy at the Census Bureau and mentioned that the type of mechanisms supported by IDP ``may be sensible'' as an alternative to differential privacy, although they worried that the relaxations might still not provide enough utility \cite[Appendix C1]{hotzetal}.
 
The reaction from the differential privacy community is also not clear. For example, IDP satisfies important criteria for formal privacy definitions such as composition and post-processing invariance \cite{idp} and a comparative survey of differential privacy variations \cite{sokdps}, written by experts, did not mention the weaknesses that we show here (namely, database reconstruction that uses  arbitrarily low amounts of privacy loss budget).

On the other hand, IDP justifies mechanisms that differential privacy experts are generally uneasy about -- IDP allows one to use something called \emph{local sensitivity} \cite{NRS07} to determine how much noise to add to query answers (however the local sensitivity itself is never released). Such mechanisms certainly do not satisfy differential privacy, specifically because sometimes they add 0 noise to queries \cite{NRS07}. In this paper, we show that having the possibility of 0 noise, while certainly useful for reconstruction, is  not necessary for reconstruction to work.

We analyze the privacy weaknesses of IDP and related privacy definitions (i.e., privacy definitions that choose their neighbors empirically). Using mechanisms recommended for IDP \cite{idp}, we demonstrate an  attack that reconstructs the entire dataset using arbitrarily low privacy loss budget (that is, while IDP claims almost no information has been revealed, the entire dataset has actually been reconstructed). We do this using queries that have a special property -- protecting those queries with IDP reveals more precise information about the data than if no protections were used at all for those queries. This is true even if the IDP mechanisms always add noise (i.e., even when the situations with 0 noise are avoided).

We then discuss how a similar type of result can be applied to a related privacy definition called Bootstrap Differential Privacy \cite{Bootstrap}. However, in this case, rather than reconstructing the entire dataset, one can only reconstruct the distinct set of records -- that is, one can determine which records are present but not how many times they appear.

We then analyze these styles of privacy definitions theoretically to determine why they have different leakage properties. Overall, we conclude that this direction is unlikely to provide the right balance between privacy and utility in practice.

 To summarize, our contributions are the following:
 \begin{itemize}[leftmargin=0.5cm,itemsep=0cm,topsep=0.5em,parsep=0.5em]
 \item We present a practical reconstruction attack against  Individual Differential Privacy \cite{idp} (IDP) and its more challenging version named $(\epsilon_1,\dots,\epsilon_k)$-Group Differential Privacy \cite{idp} (Group IDP). The privacy loss parameter for these definitions is $\epsilon$ and we show that for any $\epsilon>0$, it is possible to reconstruct any dataset whose size is larger than $2$ (or $2k$ in the case of $(\epsilon_1,\dots,\epsilon_k)$-group differential privacy). In particular, we construct queries such that answering the queries with the noise mechanisms proposed by \cite{idp} provides \emph{more} information about the data than if the queries were always answered truthfully. 
 \item We show that the reconstruction attack can be specialized to also perform membership inference and attribute inference attacks with significantly fewer queries.
 \item We then briefly consider Bootstrap Differential Privacy \cite{Bootstrap}  and show that its preferred mechanism can also be used to leak the distinct set of records in the data, again for any privacy loss $\epsilon>0$. The fact that this information can be leaked was noted by the authors \cite{Bootstrap}, but we show that it can even be leaked using the preferred mechanisms of BDP.
  \item In order to better understand these weaknesses, we consider various ad-hoc defenses against reconstruction and show that they do not solve the fundamental problems. 
  \item We also study this style of privacy definition theoretically (we call it \emph{empirical neighbors}) and show that this privacy leakage is unavoidably built-in to the privacy definition.
 \end{itemize}
 
 The rest of this paper is structured as follows. We describe notation and present background definitions in Section \ref{sec:background}.  We present a reconstruction attack against individual differential privacy and its group-based version in Section \ref{sec:idp}, where we also explain how membership and attribute inference attacks against specific individuals can be performed. This section forms the bulk of the paper.
 We then  review bootstrap differential privacy in Section \ref{sec:bootstrap} and briefly show how similar techniques can be used to launch attacks against it.
Then we analyze the weaknesses of these types of definitions more generically in Section \ref{sec:generic}. We experimentally evaluate the reconstruction algorithm for IDP in Section \ref{sec:experiments} and discuss related work in Section \ref{sec:related}. Conclusions and future work are in Section \ref{sec:conc}.
 
 \textbf{Our code and the full version of this paper with proofs can be found at \url{https://github.com/cmla-psu/idpreconstruction}.}

\section{Background And Notation}\label{sec:background}
A dataset $\data$ is a collection $\rec_1,\dots, \rec_n$ of records, each corresponding to a distinct individual. For simplicity, we assume that the total number of records $n$ is known. Each record has attributes $\att_1,\dots, \att_m$ (e.g., $\att_1=$``income'', $\att_2=$``is student?''). The value of attribute $\att_i$ for record $\rec_j$ is denoted as $\rec_j[i]$. 
The specific dataset that has been collected by a statistical agency is denoted as $\truedata$.

We say that two datasets $\data_1$ and $\data_2$ are \emph{differential privacy neighbors} (or \emph{dp-neighbors} for short) if one can be obtained from the other by modifying the record of an individual. We use the notation $\data_1\sim\data_2$ to indicate that $\data_1$ and $\data_2$ are dp-neighbors.

A mechanism $\mech$ is a (randomized) algorithm whose input is a confidential dataset and whose goal is to produce an output that protects the confidentiality of individuals whose records are in the input dataset.

\subsection{Differential Privacy}
Differential privacy is a set of restrictions on the behavior of \emph{randomized} algorithms. Intuitively, it masks the effect of any record on the output of $\mech$ by ensuring that the output distribution of $\mech$ is relatively insensitive to changes to a record in the input, hence providing plausible deniability for the contents of a record. 

\begin{definition}[$\epsilon$-differential privacy \cite{dmns06}]
A randomized algorithm $\mech$ satisfies $\epsilon$-differential privacy ($\epsilon$-DP) if for every set S $\subseteq$ Range($\mech$) and for all pairs of dp-neighbors $D_{1}$ and $D_{2}$,
\begin{align*}
    Pr[\mech(\data_1) \in S]\leq e^\epsilon Pr[\mech(\data_2) \in S]
\end{align*}
where the probability only depends on the randomness in $\mech$. 
\end{definition}
Both IDP and BDP are variations of differential privacy, but we defer their definitions to Sections \ref{sec:idp} and \ref{sec:bootstrap}, respectively, to make them relatively self-contained, so that the definition, motivation, preferred privacy mechanisms,
and attacks are all in one place.
\subsection{Sensitivity}
In the differential privacy literature, different notions of \emph{sensitivity} are used to quantify the effect that a single record could have on the output of a function $f$ and is often used to calibrate the amount of noise that a mechanism $\mech$ might add to the output of $f$.

The first of these is \emph{global sensitivity}, defined as follows:

\begin{definition}[Global sensitivity \cite{dmns06}]\label{def:globalSens}
The global sensitivity of a (vector-valued) function $f$, denoted as $\globalsens(f)$, is the largest change in $f$ that can be achieved by modifying a record in any dataset: 
\begin{align*}
    \globalsens(f) = \sup_{\data_1\sim\data_2} ||f(\data_1) - f(\data_2)||_1
\end{align*}
where the suprememum is taken over \emph{all} pairs $\data_1, \data_2$ that are dp-neighbors of each other.
\end{definition}

Global sensitivity may overestimate the amount of noise that must be added to hide the effect of a record. For this reason, 
Nissim et al. \cite{NRS07} introduced an intermediate concept called \emph{local sensitivity}. 

\begin{definition}[Local sensitivity \cite{NRS07}]\label{def:localSens}
The local sensitivity  of a (vector-valued) function $f$ with respect to a dataset $\data$ (denoted as $\localsens(f,D)$ is defined as the largest change in $f$ that can be achieved by modifying a record in $\data$:
\begin{align*}
    \localsens(f,\data) = \sup_{\data^\prime\sim \data} ||f(\data) - f(\data^\prime)||_1
\end{align*}
where the supremum is over all datasets $\data^\prime$ that are dp-neighbors of $\data$.
Note that the global sensitivity is related to local sensitivity as follows: $\globalsens(f)=\sup_{\data}\localsens(f, \data)$.
\end{definition}

Nissim et al. \cite{NRS07} noted that local sensitivity is not compatible with $\epsilon$-differential privacy. But an upper bound of it, called \emph{smooth sensitivity} is compatible with $\epsilon$-differential privacy \cite{NRS07}.  Local sensitivity, however \emph{is} compatible IDP (see Section \ref{subsec:idpreview}). The following generalization of local sensitivity is also needed:



\begin{definition}[$k$-Local Sensitivity]\label{def:GroupSens}
The $k$-local sensitivity  of a function $f$ with respect to a dataset $\data$ (denoted by $\localsens_k(f,\data)$ is defined as the largest change in $f$ that can be achieved by modifying up to $k$ records in $\data$. Let $\neigh_k(\data)$ be the set of all datasets that can be obtained from $\data$ by modifying up to $k$ records. The formula for $k$-local sensitivity is:
\begin{align*}
    \localsens_k(f,\data) = \max_{\data^\prime\in \neigh_k(\data)} ||f(\data) - f(\data^\prime)||
\end{align*}
\end{definition}
Note when $k=1$, this is the same as local sensitivity.

\section{Reconstruction against Individual Differential Privacy}\label{sec:idp}
In this section, we present reconstruction attacks against IDP and its generalization Group IDP that  is intended to provide more privacy protections \cite{idp}. We first review these privacy definitions and recommended privacy mechanisms (Section \ref{subsec:idpreview}).
We examine the main query used for the attack in Section \ref{subsec:attackquery} that tricks the privacy mechanism into revealing private information.
Using this query, we then show how to reconstruct a single column (attribute) of a table in Section \ref{subsec:onecol}.  We explain how to extend these ideas to reconstruct the entire table (Section \ref{sec:reconstruct:full}) at arbitrarily low privacy loss budget settings. Then, we explain how the attack can be specialized to membership inference and attribute inference, using many fewer queries, in Section \ref{sec:idp:discussion}.

\subsection{A Review of IDP and Group IDP}\label{subsec:idpreview}

The fundamental idea behind IDP and Group IDP is that the plausible deniability argument provided by differential privacy only needs to be applied to the actual dataset $\truedata$ collected by a data curator and does not need to apply to every possible dataset \cite{idp}. Thus IDP  only seeks to mask the differences between $\truedata$ and any dataset that can be obtained from it by modifying a record. Meanwhile, Group IDP  seeks to mask the difference between $\truedata$ and any dataset that differs from it by up to $k$ records, for some prespecified $k$.
Since Group IDP has $k+1$ parameters named $k, \epsilon_1,\epsilon_2,\dots,\epsilon_k$, we present a two-parameter simplification of it. Any mechanism that satisfies this simplification, also satisfies the more complex original definition, so any attack on the simplification also directly works on the original definition. Formally,

\begin{definition}[$\epsilon$-IDP and $(\epsilon, k)$-Group IDP \cite{idp}]\label{def:idp}
Given a fixed data set $\truedata$, privacy loss budget $\epsilon\geq 0$, and group size $k\geq 1$, let $\neigh_k$ be the set of all datasets that can be obtained from $\truedata$ by modifying up to $k$  records. A mechanism $\mech$ satisfies $(\epsilon,k)$-Group IDP \ul{with respect to $\truedata$} if for every $\data\in \neigh_k$ and every $S\subseteq\text{range}(\mech)$,
\begin{align*}
    Pr[\mech(\data) \in S] &\leq e^\epsilon Pr[\mech(\truedata)\in S]\\ 
    Pr[\mech(\truedata)\in S] &\leq e^{\epsilon}Pr[\mech(\data) \in S]
\end{align*}
When $k=1$, we say that $\mech$ satisfies $\epsilon$-IDP with respect to $\truedata$; that is, $\epsilon$-IDP is the same as $(\epsilon,1)$-Group IDP.
\end{definition}

The parameter $k$ is the group size parameter and is particularly important to reconstruction, because our attack only works on datasets of size $\geq 2k$. This is not a particularly strong restrictions because a low value of $k$ is recommended (e.g, $k=1)$ \cite{idp}.

The parameter $\epsilon\geq 0$ is the privacy loss parameter. Large values of $\epsilon$ correspond to weaker privacy protections and small values of $\epsilon$ (close to $0$) ostensibly correspond to stronger privacy protections.

We note that the original, more complex, definition has $k$ privacy loss parameters $\epsilon_1,\dots,\epsilon_k$, but  a mechanism $\mech$ satisfying  Definition \ref{def:idp} with $\epsilon=\min_i \epsilon_i$ also satisfies that more complex definition and any reconstruction attack against Definition \ref{def:idp} is therefore also a reconstruction attack against the original definition.
This privacy definition has desirable properties that are required of formal privacy definitions:
\begin{itemize}[leftmargin=0.5cm,itemsep=0cm,topsep=0.5em,parsep=0.5em]
\item \textbf{Postprocessing invariance:} Let $\mech$ be a mechanism that satisfies $(\epsilon,k)$-Group IDP  with respect to $\truedata$ and let $\randalg$ be a postprocessing algorithm whose domain contains the range of $\mech$. Then the algorithm that first runs $\mech$ and then runs $\randalg$ on the result satisfies  $(\epsilon,k)$-Group IDP with respect to $\truedata$ for the exact same privacy parameters \cite{idp}.
\item \textbf{Composition:}
Let $\mech_1$ be a mechanism that satisfies  $(\epsilon_1,k)$-Group IDP with respect to $\truedata$ and let $\mech_2$ be a mechanism that satisfies $(\epsilon_2,k)$-Group IDP with respect to $\truedata$. The mechanism that releases the outputs of both $\mech_1$ and $\mech_2$ satisfies  $(\epsilon_1+\epsilon_2,k)$-Group IDP with respect to $\truedata$.
\cite{idp}.
\end{itemize}

Mechanisms for Group IDP are based on local and $k$-local sensitivity (Definition \ref{def:GroupSens}). Specifically, the scale of the noise added to a query is proportional to the $k$-local sensitivity. Nissim et al. \cite{NRS07} earlier had argued that basing the amount of noise on local sensitivity is problematic because ``the noise magnitude itself reveals information about the database.'' They illustrated this with an example with the median function, which can have local sensitivity of 0 for some (but not all) datasets, which would result in 0 noise being added for those datasets. Their warning has often been interpreted as a caution against releasing the value of the local sensitivity \cite{hotzetal,Chetty_Friedman_2019}.

However, we demonstrate a more severe vulnerability. First, this is not a problem that affects just \emph{some} datasets -- it affects \emph{all} datasets. Second, even if the noise scale is never 0 (for example, if the noise scale is proportional to $k$-local sensitivity $+1$) and even if the local sensitivity is never revealed directly, one can still infer enough information about the dataset to reconstruct it, as long as the dataset size is $\geq 2k$.


One generic mechanism for Group IDP is  the \emph{$k$-Laplace} mechanism, defined as follows.

\begin{definition}[$k$-Laplace Mechanism \cite{idp}]
Let $g$ be a vector-valued function with $k$-local sensitivity $\localsens_k(g, \truedata)$ with respect to the true data $\truedata$. Let $\epsilon^*\in (0,\epsilon]$ be the amount of the privacy loss budget allocated to the mechanism. The $k$-Laplace mechanism outputs $g(\truedata) + \text{Laplace}(\localsens_k(g, \truedata)/\epsilon^*)$, where\\ $\text{Laplace}(\localsens_k(g, \truedata)/\epsilon^*)$ is a vector of independent Laplace random variables, each having density function $$f(x)=\frac{\epsilon^*}{2\localsens_k(g, \truedata)}\exp\left(-\frac{\epsilon^*}{2\localsens_k(g, \truedata)}|x|\right)$$
and variance $\frac{2\localsens_k(g, \truedata)^2}{(\epsilon^*)^2}$.
\end{definition}
The $k$-Laplace mechanism satisfies $(\epsilon^*,k)$-Group IDP \cite{idp} and our reconstruction attack will take advantage of the $k$-Laplace mechanism when applied to the $g$ function corresponding to the query described in Section \ref{subsec:attackquery}.

\subsection{The Attack Query}\label{subsec:attackquery}

We now identify a class of queries such that answering these queries with the $k$-Laplace mechanism and tiny values of $\epsilon^*$ (corresponding to very strong claims of privacy) ends up revealing more information about the data than if the queries were always answered truthfully (i.e., without any protections).

The queries we are interested in are \emph{predicate count queries with thresholds}. That is, given a predicate $\phi$ (a function whose input is a record and whose output is True/False) and a threshold $b$, the query $\pquery$ returns 1 if the number of records satisfying the predicate is larger than $b$. Formally,
\begin{align}
\pquery(\data) &= 
\begin{cases}
1 & \text{ if }~ \Big|\{ \rec\in\data ~:~\phi(\rec)=\text{True}\}\Big| > b\\
0 & \text{ otherwise}
\end{cases}
\label{eqn:fnv}
\end{align}

%

The $k$ local sensitivity of $\pquery$ is the following.

\begin{theoremEnd}[category=idp]{lem}
\label{lem:countKSensitivity}
Let  $k$ be a positive integer (e.g., the group size parameter in Group IDP) and suppose the true dataset $\truedata$ has $\geq k$ records. The $k$-local sensitivity of $\pquery$ with respect to $\truedata$ is 0 whenever $b<0$, $b\geq n$ (number of records in $\truedata$), or $\phi$ is always true or always false. Otherwise:
\begin{align*}
\localsens_k(\pquery,\truedata) &=
\begin{cases}
0 & \text{when } \Big|\{\rec\in \truedata~:~\phi(\rec)=True\}\Big| > b+k\\
0 & \text{when }\Big|\{\rec\in \truedata~:~\phi(\rec)=True\}\Big| \leq b-k\\
1 & \text{otherwise }
\end{cases}
\end{align*}
%
\end{theoremEnd}
\begin{proofEnd}[proof end]
The case when $b< 0$ or $b\geq n$ or $\phi$ is always true or is always false is trivial. So for the rest of the proof, we assume that none of those hold.

Let $\neigh_k$ be the set of records that can be obtained from $\truedata$ by modifying at most $k$ records.

\noindent\textbf{Case 1:} \textbf{If} $\mathbf{\Big|\{\rec\in \truedata~:~\phi(\rec)=True\}\Big| > b+k}$ then changing up to $k$ records of $\truedata$ can decrease the count by at most $k$ (i.e., by taking up to $k$ records that satisfy the predicate and changing them to some value that does not) and so for all $\data^*\in \neigh_k$, $\Big|\{\rec\in \data^*~:~\phi(\rec)=True\}\Big| > b$ and so $\pquery$ would return the same answer for $\truedata$ and for each dataset in $\neigh_k$. Thus in this case, the $k$-local sensitivity is $0$.

\noindent\textbf{Case 2:} \textbf{If} $\mathbf{\Big|\{\rec\in \truedata~:~\phi(\rec)=True\}\Big| \leq  b-k}$ then changing up to $k$ records of $\truedata$ can \emph{increase} the count by at most $k$ (i.e., by taking up to $k$ records that do not satisfy $\phi$ and changing them to a value that does). So, for all $\data^*\in \neigh_k$, $\Big|\{\rec\in \data^*~:~\phi(\rec)=True\}\Big| \leq b$ and so $\pquery$ would return the same answer for $\truedata$ and for each dataset in $\neigh_k$. Thus in this case, the $k$-local sensitivity is also $0$.

\noindent\textbf{Case 3: If}
$\mathbf{b+k\geq\mathbf{\Big|\{\rec\in \truedata~:~\phi(\rec)=True\}\Big| >  b}}$. Here we have two sub-cases:
\begin{itemize}
\item If $\Big|\{\rec\in \truedata~:~\phi(\rec)=True\}\Big|\geq k$ then one can change $k$ records that satisfy $\phi$ to a value that does not to get a  $\data^*\in \neigh_k$ for which $\Big|\{\rec\in \data^*~:~\phi(\rec)=True\}\Big| \leq  b$ (which is a decrease from the upper bound $b+k$ that defines Case 3). Thus $\pquery(\truedata)=1$ while $\pquery(\data^*)=0$ and thus the $k$-local sensitivity would be 1.
\item If $\Big|\{\rec\in \truedata~:~\phi(\rec)=True\}\Big|< k$ then,  one can modify all the records that satisfy $\phi$ to values that do not to get a $\data^*\in \neigh_k$ for which $\Big|\{\rec\in \data^*~:~\phi(\rec)=True\}\Big| =0 \leq b$ (recall that the situation where $b$ is negative has already been dealt with). Thus $\pquery(\truedata)=1$ while $\pquery(\data^*)=0$ and thus the $k$-local sensitivity would be 1.
 \end{itemize}

\noindent\textbf{Case 4: If}
$\mathbf{b\geq\mathbf{\Big|\{\rec\in \truedata~:~\phi(\rec)=True\}\Big| >  b-k}}$. Here again we have two cases:
\begin{itemize}
\item If $\Big|\{\rec\in \truedata~:~\phi(\rec)=True\}\Big|\leq n-k$ then there are at least $k$ records not satisfying $\phi$, and so $k$ of them can be modified to values that do satisfy $\phi$ to get a $\data^*\in \neigh_k$. This will increase the count by $k$ and so we will have $\Big|\{\rec\in \data^*~:~\phi(\rec)=True\}\Big| > b$. Thus $\pquery(\truedata)=0$ while $\pquery(\data^*)=1$ and thus the $k$-local sensitivity would be 1.
\item If $\Big|\{\rec\in \truedata~:~\phi(\rec)=True\}\Big|> n-k$ then there are fewer then $k$ records that don't satisfy $\phi$. If we modify all of them to have values that do satisfy $\phi$ then we get a $\data^*\in\neigh_k$ such that $\Big|\{\rec\in \data^*~:~\phi(\rec)=True\}\Big|=n>b$ (recall the situation where $b\geq n$ has already been dealth with). Thus $\pquery(\truedata)=0$ while $\pquery(\data^*)=1$ and thus the $k$-local sensitivity would be 1.
\end{itemize}

\end{proofEnd}

We are particularly interested in the queries where the predicate $\phi$ specifies a range $[u, v)$ on an attribute $\att_i$. That is $\phi(\rec)=$True if and only if $u\leq \rec[i] < v$. When $\phi$ is such a predicate, we denote the corresponding query as $\tquery$. It returns 1 when the count of records having attribute $\att_i$ in the range $[u,v)$ is larger than $b$. We call this a \emph{threshold range-count query}.

Using the $k$-Laplace Mechanism with a portion $\epsilon^*$ of the privacy budget to protect $\tquery$ results in what we shall call the Group IDP \emph{threshold range query mechanism} for $\tquery$:

\begin{eqnarray}
\mech(\data) = \tquery(\data) + \text{Laplace}\left(\frac{\localsens(\tquery,\data)}{\epsilon^*}\right)\label{eqn:idpmech}
\end{eqnarray}

Note that for some combinations of $b$ and $\data$, the $k$-local sensitivity is 0 and no noise is added. For other values of $b$ and $\data$, the local sensitivity is 1 and Laplace$(1/\epsilon^*)$ noise is added. Being able to distinguish between the two cases using only the output of $\mech$ is the key to the attack. We explain how to do this next, but we also note that having 0 noise is not necessary for the attack to work -- for example if the noise is either Laplace$(a/\epsilon^*)$ or Laplace$(b/\epsilon^*)$ for some positive numbers $a$ and $b$, reconstruction is still possible (we explain how to deal with this complication in Section \ref{sec:generic}).


\subsubsection{Detecting Noiseless Answers}
When the share of the privacy budget $\epsilon^*$ is extremely small, it is possible to detect with near perfect accuracy whether $\mech$ returned a value that has no noise ($k$-local sensitivity is 0) or is noisy ($k$-local sensitivity is 1).
For example, suppose the share of the privacy loss budget used in the mechanism is $\epsilon^*=10^{-10}$. When the $k$-local sensitivity is 1, the Laplace noise will  be a non-integer -- the probability that a floating point implementation of Laplace noise with scale $1/\epsilon^*$ is a non-integer is essentially 1. If no noise is added, then the output would certainly be 0 or 1. Thus the following decision rule has near perfect accuracy: if the output is not 0 or 1, it was because noise was added and so the local sensitivity is 1; if the output is 0 or 1, then with overwhelming probability no noise was added and local sensitivity is 0.

Moreover, even if one performs ad-hoc protections like rounding the output of the mechanism, it is still possible to tell whether the $k$-local sensitivity was 0 or 1 as follows:

\begin{itemize}[leftmargin=0.5cm,itemsep=0cm,topsep=0.5em,parsep=0.5em]
\item If  the output is rounded to the nearest integer, then if noise is injected, the probability that the output is 0 or 1 is $\leq 1-e^{-2\epsilon^*}$ (this is the probability that the absolute value of the noise is not greater than 2). When $\epsilon^* = 10^{-10}$, this probability is at most $2\times 10^{-10}$. This means that the decision rule described above will fail with probability less than $2\times 10^{-10}$.

\item If the output of the mechanism is rounded to 0 or 1 (whichever is closer to the noisy value that was produced by the mechanism), one can still distinguish between the $k$-local $\text{sensitivity}=0$ and $k$-local $\text{sensitivity}= 1$ cases. Simply ask the same query 15 times, each time with privacy loss budget share $\epsilon^*=10^{-10}/15$. The decision rule to use is: if all the 15 answers are identical then assume no noise was added and if at least 1 answer is different from the rest, then assume noise was added. Clearly, if the $k$-local sensitivity is 0 then all the 15 answers are noise-free, and the rule would be correct. On the other hand, if the $k$-local sensitivity is 1, then the probability of getting 15 ones or 15 zeroes as the answers is  approximately $2*2^{-15}\leq 10^{-10}$ and so the probability of the decision rule failing is virtually 0.  Meanwhile, the total privacy budget spent by the 15 queries is $15 * 10^{-10}/15=10^{-10}$.
\end{itemize}


\subsubsection{What one learns from noisy and noiseless answers}\label{subsub:learn}

It turns out that the ability to detect whether an answer is noisy or not allows us to infer  deterministic information about the data even if the answer was highly noisy. 
More surprisingly, answering $\tquery$ using the $k$-Laplace mechanism provides \emph{more}  information than one would get if no protection was used as all (i.e., if it was always answered truthfully no matter what). This finding is a consequence of the following lemma.

\begin{theoremEnd}[category=idp]{lem}
\label{lem:countObservation}
Let $\truedata$ be a dataset with $n$ records (where $n$ is publicly known). Let $\att_i$ be an ordered attribute and $[u,v)$ be a range that does not contain the entire domain of $\att_i$. Let $b$ be an integer threshold such that $1\leq b\leq n-1$. Let $\mech$ be the $k$-Laplace mechanism for answering the threshold range query $\tquery$. If the output $\outp$ of $\mech(\truedata)$ is released, then the following can be learned about $\truedata$:
\begin{itemize}[leftmargin=0.5cm,itemsep=0cm,topsep=0.5em,parsep=0.5em]
\item If $\outp$ is detected as a noisy output then the quantity $\Big|\{u\leq \rec[i] < v~:~\rec\in\truedata\}\Big|$  is $\geq b-k+1$ and is also $\leq b+k$. In other words, we get an upper and lower bound on the number of people in $\truedata$ whose value for $\att_i$ is in the range $[u, v)$. 
\item If $\outp$ is detected as non-noisy and $\outp=1$ then $\Big|\{u\leq \rec[i] < v~:~\rec\in\truedata\}\Big|>b+k$.
\item If $\outp$ is detected as non-noisy and $\outp=0$ then $\Big|\{u\leq \rec[i] < v~:~\rec\in\truedata\}\Big|\leq b-k$
\end{itemize}
%
\end{theoremEnd}
\begin{proofEnd}
Since the query $\tquery$ has $k$-local sensitivity either $0$ or $1$, when $\outp$ is detected as noisy it means that the $k$-local sensitivity is 1. By Lemma \ref{lem:countKSensitivity}, the $k$-local sensitivity is 1 only when the number of people in the range is $\leq b-k$ and $> b-k$. Since counts of people, $b$, and $k$ are all integers, the condition $>b-k$ is the same as $\geq b-k+1$. Hence the first item follows.

When the query answer $\outp$ is detected as non-noisy and $\outp=1$ then we learn that the number of people in the range $[u, v)$ is $>b$ (since we know we are getting the true answer). However, this means the $k$-local sensitivity is $0$ and we also know, By Lemma \ref{lem:countKSensitivity}, that this only happens when the count of people in the range is $>b+k$ or $\leq b-k$. Combined with the knowledge that it is $>b$, we have that the count of people in this range is $>b+k$. This proves the second item.

Similarly, when the query answer $\outp$ is detected as non-noisy and $\outp=0$ then we learn that the number of people in the range $[u, v)$ is $\leq b$ (since we know we are getting the true answer). However, this means the $k$-local sensitivity is $0$ and we also know, By Lemma \ref{lem:countKSensitivity}, that this only happens when the count of people in the range is $>b+k$ or $\leq b-k$. Combined with the knowledge that it is $\leq b$, we have that the count of people in this range is $\leq b-k$. This proves the third item.

\end{proofEnd}

Since our decision rule has near-perfect accuracy and uses up at most $\epsilon^*$ of the privacy loss budget (the attack would be using $\epsilon^*\leq 10^{-10}$) then we essentially know if the answer was noisy or not, and so: (1) if the answer is noisy,  we learn that something about the count of people whose attribute $A_i$ is in the range $[u,v)$. Specifically, we learn that this count is actually somewhere between $b-k+1$ and $b+k$ (an interval of size $2k-1$). \textbf{Note that answering $\tquery$ with no protection would never result in us learning that the true answer is inside such an interval};  (2) if the answer is not noisy (i.e., suppose the answer is $1$), then this non-noisy query answer directly tells us that the count of people in the range $[u,v)$ is more than $b$. But furthermore, since we have figured out that the $k$-local sensitivity is $0$, we can combine this information with Lemma \ref{lem:countKSensitivity} to learn that the count is not just $>b$, but it is in fact $>b+k$. Again, this is more information than if the query had always been answered truthfully. The reason we get so much extra information  from this $k$-local Laplace Mechanism compared to a mechanism that is always truthful, is the extra leakage caused by inferring what the local sensitivity is.  

\subsection{Single-Attribute Reconstruction.}\label{subsec:onecol}
We next show how to reconstruct one attribute $\att_i$ (one column) of the table when the data size is $\geq 2k$.\footnote{ 
We assume that the data size is public because the total number of records is a query that has a $k$-local sensitivity of 0.} That is, for  each possible value $a_j$, we will determine how many records  $\rec\in\truedata$ have $\rec[i]=a_j$. 
We consider the case where $\att_i$ is a numeric attribute since this is the hardest case. The categorical case can be handled in many ways; the simplest being to  assign an arbitrary ordering on the domain of a categorical attribute.\footnote{This is often done in practice. For example, gender is frequently coded as 0 for female, 1 for male, etc.} The amount of privacy loss budget used in this reconstruction can be made arbitrarily small. We first illustrate the attack with an example.
 


\begin{example}\label{ex:idprecon}
Let us consider $(\epsilon,1)$-Group IDP (i.e., $k=1$). Let us set the overall privacy budget at $\epsilon=0.01$. We will require each call to the threshold range query mechanism to use $\epsilon^*=10^{-10}$ of the privacy loss budget and so our goal is
to make sure that the total budget used by all the mechanism calls is at most $\epsilon=0.01$. Suppose the true dataset $\truedata$ has an income column $\att_1$, and contains 6 people whose incomes are $\{5,8,15,16,17,18\}$. An attacker can proceed as follows.
\begin{enumerate}[leftmargin=0.5cm,itemsep=0cm,topsep=0.5em,parsep=0.5em]
\item The attacker first tries to find out if, say, there are more than 3 people with incomes in the range $[1, 10)$. This means $u=1, v=10, b=3$ (and recall $k=1$). Since there are actually 2 people in that range and $2\leq b-k$, then Lemma \ref{lem:countKSensitivity} says that the $k$-local sensitivity is 0. This means that the threshold-query mechanism, even when given only $10^{-10}$ of the privacy loss budget, will output the true answer $0$. The attacker realizes that this is almost certainly not a noisy answer. Using Lemma \ref{lem:countObservation}, the attacker determines that the count of people with income in the range $[1, 10)$ is \ul{at most $b-k\equiv 2$}. 
\item The attacker can then  query if there is more than 2 people with incomes in the range $[1, 10)$. Based on the previous item, the attacker knows that there are not, but by posing this query the attacker can extract more information out of the mechanism.
So now the attacker chooses $u=1, v=10, b=2$ for the query (and recall $k=1$). Since there are 2 people in the range $[u,v)$ and $2 \not> b+k$ and $2\not\leq b-k$, then Lemma \ref{lem:countKSensitivity} says that the $k$-local sensitivity is 1. Thus  the mechanism (using $10^{-10}$ of the privacy loss budget) adds significant amounts of noise and produces an output like  $9450462192.887615$, which the attacker detects as a noisy answer. Using Lemma \ref{lem:countObservation}, the attacker determines that the number of people with income in the range $[1, 10)$ is \ul{at least $b+k-1\equiv 2$ and at most $b+k\equiv 3$}.
\item Putting the results of the previous two items together, the attacker concludes  there are at exactly 2 people with incomes in the range $[1, 10)$, and only   $10^{-10} + 10^{-10}$ privacy budget was spent on those two queries.
\item The attacker can now perform the same kind of attack on the ranges $[1, 5)$, $[5, 10)$, and $[10, \infty)$ to determine the number of people in these ranges and could keep going, subdividing the ranges until a pre-specified precision such as 1 cent -- i.e., an interval would look like $[\$9.83, \$9.84)$. Clearly, at this point the attacker would know exactly all of the incomes and as long as the attacker interacts with the mechanism less than $10^8$ times, the total privacy loss will be less than the overall target of $\epsilon=0.01$ given at the beginning of the example. Clearly, if the attacker spends even less than $10^{-10}$ privacy budget per query, the total privacy cost, according to Group IDP accounting, could be made arbitrarily small.
\end{enumerate}
%
\end{example}

\begin{figure}
\includegraphics[scale=0.5]{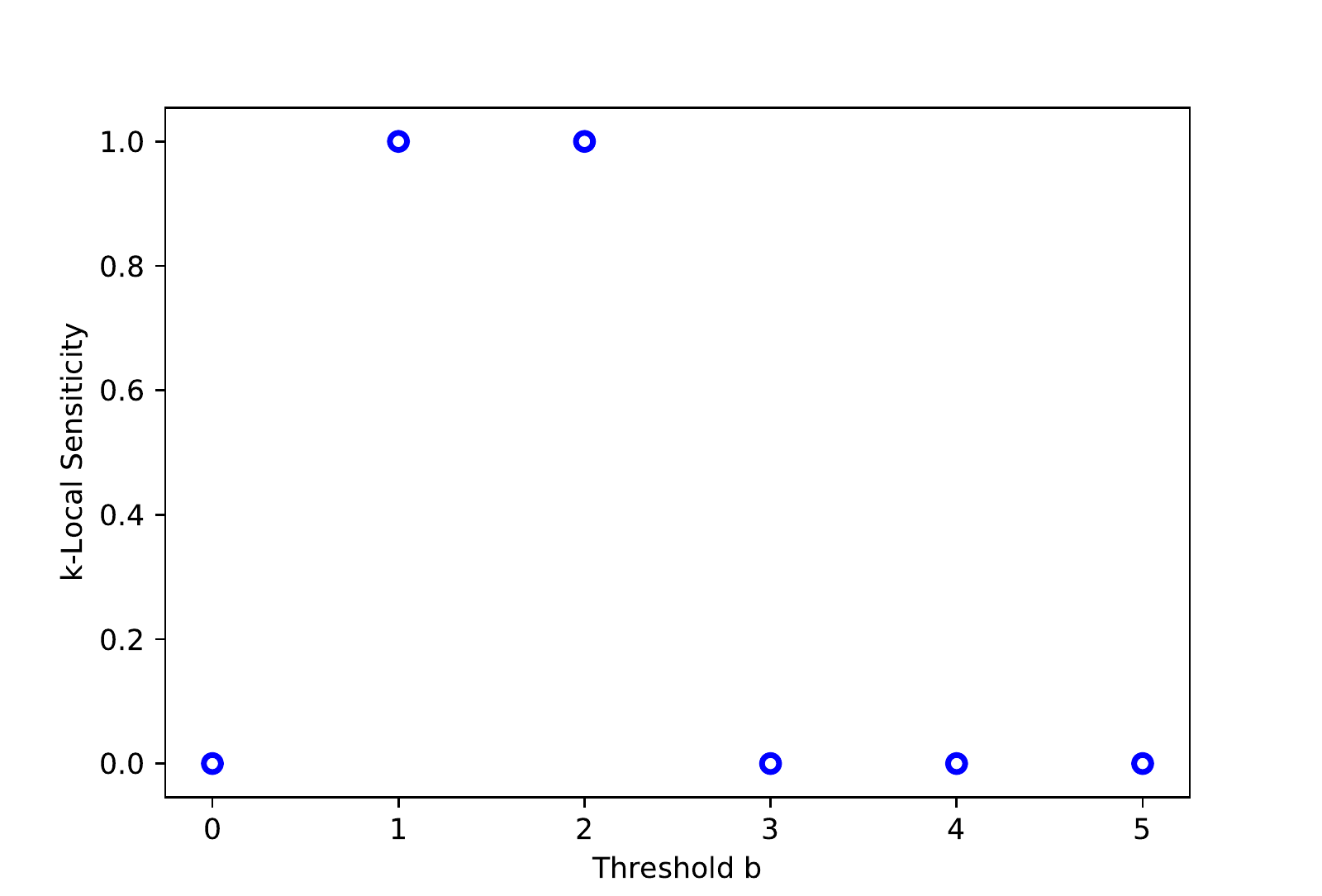}
\caption{$k$-Local sensitivity of $\tquery$ for Example \ref{ex:idprecon} as a function of the threshold $b$.}\label{fig:sens2}
\end{figure}

Thus the main subgoal for the attacker is to find out \emph{exactly} how many people have values of attribute $A_i$ in a range $[u,v)$. The attacker found a $b$ value that is at the \ul{boundary} of where the $k$-local sensitivity changes from 0 to 1 and used it to infer the true count. Indeed, as $b$ varies, the $k$-local sensitivity looks like Figure \ref{fig:sens2} -- for small $b$ the $k$-local sensitivity is 0 and the mechanism produces 1 as the noise-free answer. At some point, the $k$-local sensitivity switches to 1, and then back to $0$, after which the mechanism produces $0$ as the noise-free answer.

The following Lemma shows that this is indeed the behavior, and when one identifies a $b$ value that is on either of those two boundaries, the exact count is revealed.

\begin{theoremEnd}[category=idp]{lem}
\label{lem:boundary}
Given a predicate $\phi$, if for some integer $b^{\uparrow}$ we have (1) the $k$-local sensitivity of $\pquery[\phi][b^{\uparrow}]$ with respect to $\truedata$ is 0 and (2) the $k$-local sensitivity of $\pquery[\phi][(b^{\uparrow}-1)]$ is 1, then 
\begin{itemize}[leftmargin=0.5cm,itemsep=0cm,topsep=0.5em,parsep=0.5em]
\item The count of people in $\truedata$ whose records satisfy $\phi$ is $b^\uparrow-k$.
\item The $k$-laplace mechanism $\pquery$ will return the non-noisy answer 0 for all $b \geq b^{\uparrow}$
\end{itemize}
Furthermore, if for some integer $b^{\downarrow}$ we have (1) the $k$-local sensitivity of $\pquery[\phi][b^{\downarrow}]$ with respect to $\truedata$ is 0 and (2) The $k$-local sensitivity of $\pquery[\phi][(b^{\downarrow}+1)]$ is 1, then
\begin{itemize}[leftmargin=0.5cm,itemsep=0cm,topsep=0.5em,parsep=0.5em]
\item The count of people in $\truedata$ whose records satisfy $\phi$ is $b^\downarrow+k+1$.
\item The $k$-laplace mechanism for $\pquery$ will return the non-noisy answer 1 for all $b \leq b^{\downarrow}$
\end{itemize}
\end{theoremEnd}
\begin{proofEnd}
First, by Lemma \ref{lem:countKSensitivity}, the $k$-local sensitivity with respect to $\truedata$ changes from 0 to 1 when replacing $b^\uparrow$ with $b^\uparrow-1$ only when $\Big|\{\rec\in\truedata~:~\phi(\rec)=True\}\Big|=b^\uparrow-k$ (since the other condition for having 0 sensitivity remains unchanged as the threshold is decreased). Thus for any $b\geq b^\uparrow$, the sensitivity remains at 0 and the true answer to the query is also 0. This proves the first part.

For the second part, by Lemma \ref{lem:countKSensitivity}, the $k$-local sensitivity with respect to $\truedata$ changes from 0 to 1 when replacing $b^\downarrow$ with $b^\downarrow+1$ only when $\Big|\{\rec\in\truedata~:~\phi(\rec)=True\}\Big|=b^\downarrow+k+1$ (since the other condition for having 0 sensitivity remains unchanged as the threshold increases). Thus for any $b \leq b^\downarrow$ the sensitivity remains at 0 and the true query answer is 1. This proves the second part.

\end{proofEnd}
When applied to $\tquery$, Lemma \ref{lem:countKSensitivity} tells us that the  $k$-local sensitivity is 1 for those values of $b$ that are between  $\Big|\{u\leq \rec[i] < v~:~\rec\in\truedata\}\Big|-k$ and $\Big|\{u\leq \rec[i] < v~:~\rec\in\truedata\}\Big|+k-1$. This range contains $2k$ integers, and so if the dataset size $|\truedata|$ is $\geq 2k+1$, a boundary between $k$-local sensitivity of 0 and 1 will always exist for some $b$. Furthermore, if $|\truedata|=2k$, a boundary might not exist, but that can only happen if the count $\Big|\{u\leq \rec[i] < v~:~\rec\in\truedata\}\Big|$ is $k$. Thus, as long as $\truedata|\geq 2k$ the attacker can determine the count $\Big|\{u\leq \rec[i] < v~:~\rec\in\truedata\}\Big|$ with near perfect accuracy simply by increasing $b$ from $0$ to $|\truedata|-1$ until a boundary is found or $b$ hits its upper limit. As long as the attacker splits his privacy budget across these (at most) $|\truedata|$ queries, he can reconstruct the true count almost perfectly at arbitrarily low ``privacy cost.'' The pseudocode is shown in Algorithm \ref{alg:localCount}. For simplicity, it shows the linear search but we note that a binary search can be used instead.



\begin{algorithm}
\caption{Reconstructing counts of records with attribute $\att_i$ in an interval $[u,v)$. }\label{alg:localCount}
\DontPrintSemicolon
\SetKwProg{Function}{def}{:}{}
$k,\epsilon$: Group IDP parameters\;
$n\geq 2k$: publicly known number of records in $\truedata$\;
$i$: index of the target attribute\;
$\mquery$: mechanism that answers $\tquery$ using the $k$-Local Laplace mechanism as in Equation \ref{eqn:idpmech}\;
\Function{CountReconstruct($k$, $\epsilon_{target}, u, v, i$)}{
$b_0\gets 0$\;
$a_0\gets$  result of $\mquery[i][b_0][u][v]$ using privacy  budget $\frac{\epsilon}{n}$\;
\tcp{Note linear search is shown for simplicity}
\tcp{Use binary search for more efficiency}
\For{$j=1,\dots, n-1$}{
$b_j\gets j$\;
$a_j\gets$ result of $\mquery[i][b_j][u][v]$ using privacy  budget $\frac{\epsilon}{n}$\;
  \uIf{$a_j$  detected as noisy, $a_{j-1}$ detected as non-noisy}{
      \Return $b_{j-1}+k+1$
  }
  \ElseIf{$a_j$  detected as non-noisy, $a_{j-1}$ detected as noisy}{
     \Return $b_j-k$
  }
}
\tcp{After loop, either all answers were noisy}
\tcp{or all answers were non-noisy}
\tcp{But all non-noisy is impossible}
\Return $k$
}
\end{algorithm}

\begin{algorithm}
\caption{Reconstructing all elements in the column corresponding to attribute $\att_i$}\label{alg:columnreconstruct}
\DontPrintSemicolon
\SetKwProg{Function}{def}{:}{}
$k,\epsilon$: Group IDP parameters\;
$\gamma$: targeted decimal point precision of each reconstructed element\;
$n\geq 2k$: publicly known data size\;
\Function{ColumnReconstruct($k$, $\epsilon_{target}$, $\gamma$)}{
$u\gets $ Lower bound on domain of $\att_i$\;
$v\gets $ Upper bound on domain of $\att_i$\;
\tcc{Target privacy parameter for each call to Algorithm \ref{alg:localCount}: CountReconstruct()}
$\epsilon_{share} = \min(10^{-10},  \frac{\epsilon}{(v-u)/\gamma})$\;
$vals\gets []$\tcp*{Will store reconstructed values}
$s_0\gets n$ \tcp*{Number of items left to reconstruct}
\tcp{Note linear search is shown for simplicity}
\tcp{Use binary search for more efficiency}
\While{$s_0\neq 0$}{
   $v \gets v-\gamma$ \tcp{decrease upper bound}
$s_1\gets$ CountReconstruct($k$, $\epsilon_{share}$, u, v, i)\;
  \uIf{$s_0\neq s_1$}{
  \tcp{There are $s_0-s_1$ items in  $[v, v+\gamma)$}
  add $s_0-s_1$ copies of $v$ into $vals$ array\;
  $s_0\gets s_1$\;


  }


}

\Return $vals$
    


    
   

    
}
\end{algorithm}
Now that we have a tool for determining the counts of records within a range $[u,v)$, we can use it to reconstruct an entire attribute $A_i$ up to a certain precision -- that is, we can find all of the incomes in the dataset up to the nearest cent.  Algorithm \ref{alg:columnreconstruct} shows how to do this. The algorithm starts by setting $u$ to be the lower bound on the domain of Attribute $\att_i$ and $v$ to be an upper bound.\footnote{If bounds are not not known in advance, one could start with the interval $[-1, 1)$ and keep doubling the endpoints as long as Algorithm \ref{alg:localCount} reports that less than $n$ records are in the interval.} For example for the Income attribute, one could set $u=0$ and $v=2^{40}$. 

Next, the algorithm considers a decreasing sequence of values $v=v_0 > v_1 > v_2 > \dots$. It calls Algorithm \ref{alg:localCount} to find out how many people have attribute $\att_i$ in the range $[u, v_j)$. If it finds that the count for $[u, v_j)$ (call this count $s_0$) and the count for $[u, v_{j-1})$ (call it $s_1$) are different, then there must be $s_0-s_1$ people in the range $[v_{j-1}, v_j)$. If the width of the interval is $\leq 0.01$, then it has reconstructed those income values up to a penny. In general, the target precision is a user-provided input called $\gamma$.

Note that Algorithm \ref{alg:columnreconstruct} uses linear search to find the next value after $v_j$ for which the count changes, but this is shown for simplicity and can be replaced by a binary search instead for efficiency.

Algorithm \ref{alg:columnreconstruct} also upper bounds the number of calls it would need to Algorithm \ref{alg:localCount} and splits its target  privacy budget $\epsilon$ equally among these calls.  It ensures that the amount of privacy budget given to each Algorithm \ref{alg:localCount} call is small enough so that an answer to a $k$-local Laplace mechanism can be detected as noisy/non-noisy  and so that Algorithm \ref{alg:columnreconstruct} can meet its budget goals.



\subsection{Reconstructing the Full Dataset}\label{sec:reconstruct:full}

Reconstructing the full dataset can be done in an iterative manner. One first reconstructs the first attribute $\att_1$ using Algorithm \ref{alg:columnreconstruct}. This gives a set of records $\rec_1,\dots,\rec_n$ that have just one attribute. One then needs to add attribute $\att_2$ to each record, then attribute $\att_3$, and so on.
%
Since the algorithms used to do this are nearly identical to Algorithms \ref{alg:localCount} and \ref{alg:columnreconstruct}, we do not list them here, but instead explain how the process would work with an example.

\begin{example}
Suppose Algorithm \ref{alg:columnreconstruct} has been used to reconstruct the Age column to get the values $[18, 18, 21, 21, 30]$. 
To add the next column, say height, we would be interested in queries of the form: ``are there more than $b$ many 18-year-olds who have \emph{height} in $[u, v)$.''  This is another predicate count query with a threshold $b$ and its $k$-local sensitivity is again given by Lemma \ref{lem:countKSensitivity}. It is answerable using the $k$-local Laplace mechanism, similar to  Equation \ref{eqn:idpmech}. To identify the number of 18-year-olds who have height in $[u,v)$, again one would search for a $b$ value on the boundary of $k$-local sensitivity changes, using Algorithm \ref{alg:localCount}, but modified to use the $k$-local laplace mechanism for this new query. Then using Algorithm \ref{alg:columnreconstruct} with this modified Algorithm \ref{alg:localCount} allows us to find all of the heights associated with 18-year-olds in the data. Then we would repeat the process with 21-year-olds and 30-year-olds.

Continuing this process with a third attribute, then a fourth, etc., would result in the entire dataset being reconstructed as long as it contains at least $2k$ people.
\end{example}

\subsection{Additional Attacks}\label{sec:idp:discussion}
The attack algorithm can be made efficient by replacing linear search in Algorithms \ref{alg:localCount} and \ref{alg:columnreconstruct} with binary search.
The algorithms could also be adapted for other kinds of attacks, not just entire data reconstruction.

\begin{example}[Confirmation of Uniqueness]\label{ex:attack:unique} Suppose the dataset schema is $\att_1,\dots, \att_m$ and we know the values of $\ell$ of these attributes for a target individual's record $\rec^*$ (say we know $\rec^*[\att_1]=a_1,\dots, \rec^*[\att_{\ell}]=a_\ell$). We may ask if that person is unique in the data for those attributes. We can consider the following query $q_b$: is the number of records $\rec$ with $\rec[\att_1]=a_1,\dots, \rec[\att_{\ell}]=a_\ell$ larger than $b$? This is a predicate count query with threshold $b$ and its $k$-local sensitivity is given in Lemma \ref{lem:countKSensitivity}. Namely, when the true count of such records is $>b+k$ or $\leq b-k$, the $k$-local sensitivity is 0, otherwise it is 1. Let $\mech_b$ be the  $k$-laplace mechanism for this query. Then we run $M_b$ with $b=k$ and a tiny privacy budget and then we run it with $b=k+1$. By Lemma \ref{lem:boundary}, this would be an upper boundary for the $k$-local sensitivity change (i.e., the $k$-local sensitivity is detected as 1 for $b=k$ and detected as $0$ for $b=k+1$) if and only if  there truly is only one such person in the data. Thus if we observe this combination of non-noisy answer for $b=k+1$ and a noisy answer for $b=k$, we learn the person is unique in the dataset on those attributes. On the other hand, if we observe a different outcome then we learn that the person is not unique. This attack only requires 2 accesses to the mechanism.
\end{example}

\begin{example}[Membership Inference]\label{ex:memberinference}
Suppose we know that an individual is unique in the population based on attributes $\att_1,\dots, \att_\ell$ and we want to know whether they are in the dataset (e.g., it could be an HIV dataset). This attack would proceed in the same way as in Example \ref{ex:attack:unique}, except we run the mechanism  $\mech_b$ from that example with $b=k$ and $b=k-1$ (with very small privacy budgets). If there are 0 people in the dataset with those combinations of attributes, the $k$-local sensitivity would be $1$ when $b=k-1$ and 0 when $b=k$. Thus again, this attack looks for the upper boundary and that is why 2 mechanism calls are needed (i.e., to identify which boundary it is).

Note that if the query was never protected, we would simply ask 1 query: if the number of people with a particular combination of uniquely identifying attributes is $> 0$. If the answer is True, then the person is in the dataset, if False, then they are not. This is an interesting observation because, even though the $k$-local laplace mechanism reveals more precise information about the dataset (as explained in Section \ref{subsub:learn}), this more precise information is more complex: (1) when the query is protected, there are two boundaries and we need to determine which one we found; (2) when the query is unprotected, there is only one relevant boundary -- the $b$ at which the answer changes from True to False. If a unique person is in the data, this boundary would occur at $b=0$.
\end{example}

\begin{example}[Attribute Inference]\label{ex:att}
Suppose we know that an individual is in the dataset and we know that the values for attributes $\att_1,\dots, \att_\ell$ for that individual. We may be interested in learning the value of $\att_{\ell+1}$. This is exactly what the dataset reconstruction algorithm does (section \ref{sec:reconstruct:full}) -- it finds the multiset of values for $\att_{\ell+1}$ for the records for which $\att_1=a_1,\dots, \att_\ell=a_\ell$. The reconstruction algorithm does this for all combinations of $(\att_1,\dots, \att_\ell)$ values that appear in the dataset, but clearly it can be specialized to target just one particular combination as well.
\end{example}

\section{A Brief Examination of Bootstrap Differential Privacy}\label{sec:bootstrap}
In this section, for the purposes of comparison with IDP, we review bootstrap differential privacy (BDP) \cite{Bootstrap}  and demonstrate how it can leak the distinct set of records in the data, using the preferred mechanism in \cite{Bootstrap}. 

\subsection{Bootstrap Differential Privacy}
BDP again defines its own versions of neighbors, sensitivity, and Laplace mechanism \cite{Bootstrap} as follows.

\begin{definition}[Bootstrap Neighbors \cite{Bootstrap}]
Given a true dataset $\truedata$, we say that $\data_1$ and $\data_2$ are bootstrap neighbors conditioned on $\truedata$ if $\data_1$ can be obtained from $\data_2$ by replacing one record and both $\data_1$ and $\data_2$ can be obtained from $\truedata$ by changing the multiplicities of records in $\truedata$ (i.e., $\data_1$ and $\data_2$ cannot contain a record that $\truedata$ does not contain).
\end{definition}

\begin{definition}[$\epsilon$-bootstrap differential privacy (BDP) \cite{Bootstrap}]
Given a dataset $\truedata$ and privacy parameter $\epsilon>0$, 
a randomized algorithm $\mech$ satisfies  $\epsilon$-bootstrap differential privacy if for every set S $\subseteq$ Range($\mathcal{M}$) and for all pairs of bootstrap neighboring data sets $\data_1$ and $\data_2$ (conditioned on $\truedata$),
\begin{equation}
    Pr[\mech(\data_1) \in S]\leq e^\epsilon Pr[\mech(\data_2) \in S]
\label{eqn:DPDef}
\end{equation}
\end{definition}

The Bootstrap sensitivity (BS) of a function $f$ with respect to  $\truedata$ takes the usual definition of sensitivity from differential privacy and swaps in bootstrap neighbors conditioned on $\truedata$.

\begin{definition}[Bootstrap Sensitivity \cite{Bootstrap}]
The Bootstrap sensitivity (BS) of  $f$ with respect to  $\truedata$, denoted by $\localsens_B(f,\data)$ is
\begin{equation}
    \localsens_B(f,\truedata) = \max_{\data_1\sim \data_2} ||f(\data_1) - f(\data_2)||_1
\label{eqn:bootlocalSens}
\end{equation}
where the maximum is taken over bootstrap neighbors conditioned on $\truedata$.
\end{definition}
Bootstrap sensitivity is used to calibrate noise for the bootstrap Laplace mechanism:

\begin{definition}[Bootstrap Laplace Mechanism \cite{Bootstrap}]
Let $f$ be a function whose output is a vector. Let $\epsilon^*>0$ be a privacy parameter. The bootstrap Laplace mechanism is a mechanism $\mech$  that, on input $\truedata$, adds independent Laplace noise with scale $\localsens_B(f, \truedata)/\epsilon^*$ to each component of $f$ (i.e. $\mech(\data) = f(\data) + $Laplace$(\localsens_B(f, \data)/\epsilon^*)$. 
\end{definition}

We next show how the bootstrap Laplace mechanism can be used to verify the existence or non-existence of any record with almost no privacy expenditure. Let $\phi$ be an arbitrary predicate (e.g., ``Income > 50k and Age = 32'') and let $\query_\phi$ be the query that returns 1 if and only if some record in $\truedata$ satisfies $\phi$:
\begin{align*}
\query_\phi(\data) &=
\begin{cases}
1 & \phi(\rec)=\text{True for some }\rec\in\data\\
0 & \phi(\rec)=\text{False for all } \rec\in \data
\end{cases}
\end{align*}

\begin{theoremEnd}[category=bdp]{lem}\label{lem:bdpattack}
Given a true dataset $\truedata$, the bootstrap sensitivity of $\query_\phi$ with respect to $\truedata$ is:
\begin{align*}
\localsens_B(\query_\phi,\truedata) &=
\begin{cases}
0 & \text{ if all $\rec\in\truedata$ satisfy $\phi$}\\
0 & \text{ if no $\rec\in\truedata$ satisfies $\phi$}\\
1 & \text{otherwise}
\end{cases}
\end{align*}
\end{theoremEnd}
\begin{proofEnd}
Let $\data_1,\data_2$ be bootstrap neighbors conditioned on $\truedata$. This means that any record in $\data_1$ and $\data_2$ also appears in $\truedata$. Hence if all records in $\truedata$ give the same answer for $\phi$ (i.e., all records satisfy it or all do not) then $\query_\phi(\data_1)=\query_\phi(\data_2)$ and so the bootstrap sensitivity is 0.

If there is some record $\rec_1\in \truedata$ for which $\phi(\rec_1)=$True and a $\rec_2\in\truedata$ for which $\phi(\rec_2)=$False, then $\data_1=\{\rec_1\}$ and $\data_2=\{\rec_2\}$ are bootstrap neighbors conditioned on $\truedata$ and $\query_\phi(\data_1)-\query_\phi(\data_2)=1$, hence the bootstrap sensitivity is 1.
\end{proofEnd}

Thus, if one uses the bootstrap Laplace mechanism with a tiny privacy loss budget (e.g., $\epsilon=10^{-10}$) to answer $\query_\phi$, then Lemma \ref{lem:bdpattack} tells us that:
\begin{enumerate}
\item If we receive the answer 0, then with overwhelming probability, no record in $\truedata$ satisfies $\phi$ (because it is almost impossible for an extremely noisy answer to be 0 or 1, hence this must have been a noise-free answer and the bootstrap sensitivity would have to be 0).
\item If we receive the answer 1, then with overwhelming probability, all records in $\truedata$ satisfy $\phi$.
\item If we receive any other answer, it is definitely a noisy answer (bootstrap sensitivity is 1) and so there exists a record in $\truedata$ satisfying $\phi$.
\end{enumerate}

Thus the output of the bootstrap Laplace mechanism tells us whether there is or is not such a record in the database (i.e., we have figured out what the true answer to $\query_{\phi}$ is) and sometimes it tells us more information (when all records satisfy $\phi$ or all do not). So again, protecting the query with the bootstrap Laplace mechanism reveals at least as much information compared to always answering $\query_\phi$ accurately. It allows us to probe which records are in $\truedata$ (but not how many copies there are).

\section{A General Study of Empirical Neighbors Definitions}\label{sec:generic}
We have demonstrated an attack against $(\epsilon,k)$-Group IDP that recovers any dataset with at least $2k$ records and sketched an attack against BDP that recovers all records in the dataset but not their multiplicity. Both attacks work with arbitrarily low privacy budget parameters $\epsilon$ (which, according to those definitions should correspond to strong privacy protections).
In this section, we consider whether there are simple fixes for this style of privacy definition that can prevent reconstruction or whether the problems are deeper and harder to fix.

\subsection{Ensuring all answers are noisy}
In the previous sections, we exploited the fact that we can detect whether a query answer is noisy or not using arbitrarily small amounts of privacy budget. What if one changes the mechanism so that noise is always added?
For example, consider the following modification to the $k$-local Laplace Mechanism: add 1 to the $k$-local sensitivity and use that to calibrate the noise. That is, if $g$ is a function with $k$-local sensitivity $\localsens_k(g,\truedata)$ with respect to $\truedata$, then this modified mechanism $\mech^\dagger$ returns:
\begin{align*}
\mech^\dagger(\truedata;\epsilon) = g(\truedata) + \text{Laplace}\left(\frac{\textcolor{red}{1+}\localsens_k(g,\truedata)}{\epsilon}\right)
\end{align*}
Such a mechanism, when given privacy loss budget $\epsilon$, would be answering the threshold range-count queries $\tquery$ from Section \ref{subsec:attackquery} by either adding Laplace$(2/\epsilon)$ noise (when the the $k$-local sensitivity is 1) or Laplace$(1/\epsilon)$ noise (when the $k$-local sensitivity is 0). If we can reliably detect which type of noise was added, then the same reconstruction attacks from Section \ref{sec:idp} could be used.

It turns out that this is also possible using statistical hypothesis testing and exploiting the composition rules for privacy definitions like IDP and BDP. For example, given a desired target $\epsilon$ and an integer $m$, we can run the mechanism $\mech^\dagger$ for $m$ times, each time using $\epsilon/m$ privacy budget for a total cost of $\epsilon$. This gives us $m$ noisy numbers $z_1,\dots, z_m$ which are either obtained by adding Laplace$(2m/\epsilon)$ noise to an unknown quantity, or Laplace$(m/\epsilon)$ noise. We can use the empirically observed variance as a  test statistic $\psi_m$:

\begin{align*}
\psi_m = \frac{\epsilon^2}{m^2}\left[ \frac{1}{m-1}\sum_{i=1}^m\left(z_i - \frac{z_1+z_m}{m}\right)^2\right]
\end{align*}

If the $z_i$ are generated with Laplace$(2m/\epsilon)$ noise, then the variance of each $z_i$ is $8m^2/\epsilon^2$ and so the expected value of $\psi_m$ would be $\frac{\epsilon^2}{m^2} \frac{8m^2}{\epsilon^2}=8$. On the other hand, if the $z_i$ are generated with Laplace$(m/\epsilon)$ noise, then the variance of each $z_i$ is $2m^2/\epsilon^2$ and so the expected value of $\psi_m$ would be $\frac{\epsilon^2}{m^2} \frac{2m^2}{\epsilon^2}=2$.

Thus, there is a simple decision rule one could use. Run the mechanism $\mech^\dagger$ $m$ times, with $\epsilon/m$ privacy budget each time. Compute the test statistic $\psi_m$ and if it is $<5$, decide that Laplace$(2m/\epsilon)$ was used (hence $k$-local sensitivity is 1), otherwise decide that Laplace$(2m/\epsilon)$ (hence $k$-local sensitivity is 0). If this decision rule is highly accurate then this is all that is needed to do reconstruction using the algorithms in Section \ref{sec:idp}.

The following lemma shows that when $m$ is large enough, $\psi_m$ is highly concentrated around its mean (either 2 or 8) and so the decision rule is very accurate. An empirical demonstration is also shown in Table \ref{tab:decisionsim} which shows the empirical accuracy for different values of $m$. It is based on 2 million simulations, in which half of the simulations used Laplace$(2m/\epsilon)$ noise and the other half used Laplace$(m/\epsilon)$ noise. Note  the total privacy budget expended is always $\epsilon$, regardless of the value of $m$, and that by Lemma \ref{lem:decisionrule}, the decision rule has the same accuracy for any value of $\epsilon>0$.

\begin{table}
\begin{tabular}{|c|c|}\hline
$\mathbf{m}$ & accuracy\\\hline
10 & $\frac{1,606,049}{2,000,000}=\phantom{0}80.30245\%$\\
100 & $\frac{1,976,433}{2,000,000}=\phantom{0}98.82165\%$\\
1000 &$\frac{2,000,000}{2,000,000}=100.00000\%$\\\hline
\end{tabular}
\caption{Empirical accuracy of the decision rule based on $\phi_m$, for different values of $m$ for 2 million simulations.}\label{tab:decisionsim}
\end{table}

\begin{theoremEnd}[category=general]{lemma}\label{lem:decisionrule}
Given an integer $m>1$ and any $\epsilon>0$ and a noise scale multiplier $\alpha\geq 0$, define the following random variables:
\begin{enumerate}
\item $z_1,\dots, z_m$, where each $z_i= \mu$+Laplace$(\alpha m/\epsilon)$ for some unknown number $\mu$ (the private value that gets noised).
\item $z^{*}_1,\dots, z^*_m$ where each $z^*=$Laplace$(1)$
\end{enumerate}
Furthermore, define:
\begin{align*}
\psi_m &= \frac{\epsilon^2}{m^2}\left[ \frac{1}{m-1}\sum_{i=1}^m\left(z_i - \frac{z_1+z_m}{m}\right)^2\right]\\
\psi_m^{*} &=  \frac{1}{m-1}\sum_{i=1}^m\left(z^{*}_i - \frac{z^{*}_1+z^{*}_m}{m}\right)^2\\
\end{align*}
Then the distribution of $\psi_m$ is the same as the distribution of  $\alpha^2\psi_m^*$ (in particular, it doesn't depend on $\epsilon$ or the private value $\mu$). The expected value of $\psi_m^*$ (resp., $\psi_m$) is 2 (resp., $2\alpha^2$) and $\psi_m^*$   converges to 2  with probability 1 as $m\rightarrow\infty$ (hence $\psi_m$ converges to $2\alpha^2$).
\end{theoremEnd}
\begin{proofEnd}[proof end]
We first note that $\frac{\epsilon}{m}(z_i-\mu)$ is a Laplace$(\alpha)$ random variable (because scaling it by $\epsilon/{m}$ is the same as multiplying the scale parameter by $\epsilon/m$) so it has the same distribution as $\alpha z^*_i$. Hence
\begin{align*}
\lefteqn{
\frac{\epsilon^2}{m^2}\left[ \frac{1}{m-1}\sum_{i=1}^m\left(z_i - \frac{z_1+z_m}{m}\right)^2\right]} \\
&= \frac{\epsilon^2}{m^2}\left[ \frac{1}{m-1}\sum_{i=1}^m\left(z_i - \mu+ \mu - \frac{z_1+z_m}{m}\right)^2\right]\\
&= \frac{\epsilon^2}{m^2}\left[ \frac{1}{m-1}\sum_{i=1}^m\left((z_i - \mu)  - \frac{(z_1-\mu)+(z_m-\mu)}{m}\right)^2\right]\\
&= \left[ \frac{1}{m-1}\sum_{i=1}^m\left(\frac{\epsilon}{m}(z_i - \mu)  - \frac{\frac{\epsilon}{m}(z_1-\mu)+\frac{\epsilon}{m}(z_m-\mu)}{m}\right)^2\right]\\
\intertext{and so has the same distribution as}
&= \left[ \frac{1}{m-1}\sum_{i=1}^m\left(\alpha z^{*}_i   - \frac{\alpha z^{*}_1+\alpha z^{*}_m}{m}\right)^2\right]\\
&= \alpha^2 \left[ \frac{1}{m-1}\sum_{i=1}^m\left(z^{*}_i   - \frac{z^{*}_1+z^{*}_m}{m}\right)^2\right]\\
\end{align*}
and so $\psi_m$ has the same distribution as $\alpha^2 \psi^*_m$.

Now, the formula for $\psi^*_m$ is known as the sample variance of a sequence of iid random variables and is known to be an unbiased estimate of their variance. Since the variance of Laplace(1) is 2, the expected value of $\psi^*_m$ is 2 and the expected value of $\psi_m$ is $2\alpha^2$. By the law of large numbers, the convergence happens with probability 1.
\end{proofEnd}

Thus, the foundational mechanisms for these privacy definitions are flawed and reconstruction-proof fixes most likely require more complex strategies like smooth sensitivity \cite{NRS07} in differential privacy. We next examine flaws in the formulation of the privacy definitions themselves.

\subsection{Is leakage built in to the privacy definition?}

We saw that simple modifications to the mechanisms to make sure that they always add noise is still not sufficient to protect against reconstruction (one would need to use something much more complex, such as smooth sensitivity \cite{NRS07} and differential privacy). So next we study the general class of privacy definitions that IDP and BDP belong to in order to identify further flaws. We call this class of definitions \emph{empirical neighbors}.
The main components of empirical neighbors privacy definitions are:
\begin{enumerate}[leftmargin=0.5cm,itemsep=0cm,topsep=0.5em,parsep=0.5em]
\item A set of pairs of neighbors to protect. This set depends on $\truedata$, the data observed by the data collector. Hence we represent it as $\npairs(\truedata)$. The privacy constraints are obtained from $\npairs(\truedata)$ -- for each $(\data_1,\data_2) \in\npairs(\truedata)$ and each possible output $\outp$ of a mechanism $\mech$, they require that $P(\mech(\data_1)=\outp)\leq e^\epsilon P(\mech(\data_2)=\outp)$. For example, in IDP, $\npairs(\truedata)$ has the form $$\{(\truedata, \data_1), (\truedata,\data_2),\dots\} \cup \{(\data_1,\truedata), (\data_2, \truedata), \dots\}$$ where $\data_1,\data_2,\dots$ are the datasets that can be obtained from $\truedata$ by replacing one record. Similarly, in BDP, $\npairs(\truedata)$ contains all pairs $(\data_1,\data_2)$ where $\data_1$ can be obtained from $\data_2$ by replacing one record and all records that appear in $\data_1$ and $\data_2$ must also appear in $\truedata$.
\item A hint function $\hint$ that looks at the data. The data collector decides which mechanism to use based on the hint $\hint(\truedata)$. We note that the use of such hint functions is becoming increasingly common. Not only is it implicitly used in IDP \cite{idp} and BDP \cite{Bootstrap} but it was also used for actual data releases for the Opportunity Atlas \cite{Chetty_Friedman_2019} and the 2020 Decennial Census (the as-enumerated population count in each state as well as the number of housing units and non-empty group quarters in each geographic area) \cite{Abowd20222020}. In fact, many papers on differential privacy implicitly use $\hint(\truedata)=|\truedata|$ because they reveal the exact size of the dataset.
\item A mechanism selector $\chooser$ whose input is $\hint(\truedata)$ and whose output is a mechanism that satisfies the constraints obtained by $\npairs(\truedata)$. This reflects the core principles in IDP and BDP that a mechanism is chosen after observing the data.
\end{enumerate}

These pieces fit together into a privacy definition, generalizing IDP and BDP, as follows:
\begin{definition}[Empirical Neighbors]
Given $\npairs$ and a hint function $\hint$, a mechanism chooser satisfies $\epsilon$-empirical neighbors privacy if for any choice of $\truedata$, then  $\chooser(\hint(\truedata))$ produces a mechanism $\mech$ that satisfies:
\begin{align*}
Pr[\mech(\data_1)=\outp]\leq e^\epsilon Pr[\mech(\data_2)=\outp]
\end{align*}
for all $(\data_1,\data_2)\in \npairs(\truedata)$ and all possible outputs $\outp$.
\end{definition}

Both IDP and BDP don't explicitly state the rules that must be followed when choosing a mechanism -- what information about $\truedata$ can be used? Equivalently, what restrictions are there on the hint function $\hint$? Because these rules were not fully specified, our attacks had to use the mechanism design principles provided by those papers.

Some natural choices for $\hint$ are: (1) no restrictions, (2) $\hint(\truedata)=\emptyset$, (3) $\hint(\truedata)=\npairs(\truedata)$ -- in other words, $\hint$ provides information equivalent to the set of constraints that the mechanism selected by $\chooser$ should satisfy (this is most likely what was intended in IDP and BDP).

We next show the consequences of each of these choices, which is that this style of definition allows $\hint(\truedata)$ (the information used to decide on a mechanism) to be leaked, which can be catastrophic in the cases of IDP and BDP. Furthermore, preventing the leakage of $\hint(\truedata)$ results in differential privacy.

\begin{theoremEnd}[category=general]{lem}\label{lem:hint}
The empirical neighbors definitions allow the release of $\hint(\truedata)$ for any $\epsilon$ parameter. In particular, if $\hint(\truedata)=\npairs(\truedata)$ then Group IDP allows $\truedata$ to be revealed whenever $|\truedata|>1$ and BDP allows the distinct set of records to be revealed.

On the other hand, if $\hint(\truedata)=0$ and $\bigcup_D \npairs(\data)$ is the set of all pairs of datasets that differ on a record, then the empirical neighbors definition is equal to differential privacy.
\end{theoremEnd}
\begin{proofEnd}
For any set of bits $b$, let $\mech_{b}$ be the mechanism that ignores its input and simply outputs $b$. Consider the chooser function such that  $\chooser(\hint(\truedata))$ that returns $\mech_b$, where $b=\hint(\truedata)$. Clearly this satisfies empirical neighbors privacy for any privacy parameter $\epsilon$ and always reveals $\hint(\truedata)$.

If $\hint(\truedata)=\npairs(\truedata)$, as is the case with IDP and BDP, then we can reason as follows. For the case of IDP and Group IDP, $\npairs(\truedata)$ consists of pairs $(\data_1,\data_2)$ where either $\data_1=\truedata$ or $\data_2=\truedata$, so $\truedata$ is the dataset that appears in every pair. If $|\truedata|>1$ then there are at least 2 pairs and only $\truedata$ will appear in all of them, hence $\truedata$ is revealed.

In the case of BDP, if we take all of the rows of all of the datasets that appear in $\npairs(\truedata)$ and then apply the database distinct operator, we get the distinct rows in $\truedata$.

Finally, if $\hint(\truedata)=\emptyset$, then a mechanism $\mech$ must be chosen without looking at the data, and so letting $\mathcal{N}=\bigcup_{\data} \npairs(\data)$ be the set of all pairs of databases that are neighbors for some dataset, the condition of the lemma is that $\mathcal{N}$ covers all pairs of neighbors $(\data_1,\data_2)$ that differ on one record and so the only way to guarantee that the empirical neighbors constraints are always satisfied is to ensure they are satisfied for all $\data_1,\data_2\in\mathcal{N}$ which is equivalent to differential privacy.
\end{proofEnd}

\section{Experiments}\label{sec:experiments}
 As a proof of concept, we empirically evaluate the attacks against IDP because of its leakage potential.  We consider 3 attack scenarios:
\begin{itemize}[leftmargin=0.5cm,itemsep=0cm,topsep=0.5em,parsep=0.5em]
\item \textbf{Membership inference}: given a set of uniquely identifying attributes of a target individual, how many queries does an attacker need to verify that the target individual is in the dataset, using arbitrarily low privacy budget.
\item \textbf{Attribute inference}: given a set of uniquely identifying attributes of a target individual, how many queries does an attacker need to reconstruct the rest of the target's record, using arbitrarily low privacy budget.
\item \textbf{Full dataset reconstruction}: how many queries does an attacker need to reconstruct the entire dataset, using arbitrarily low privacy budget.
\end{itemize}
In these experiments, we optimize the attack code of Algorithms \ref{alg:localCount} and \ref{alg:columnreconstruct} to use binary search instead of sequential search. We  compare (1) how many queries are needed when they are ``protected'' by the $k$-local Laplace mechanism vs. (2) how many queries are needed when no protection is used (i.e., they are always answered without noise). When reconstructing using ``protected'' threshold range-count queries $\tquery$ the main idea is to look for a threshold $b$ for which the threshold range query mechanism $\mech$ switches from noisy to non-noisy answers. When reconstructing using unprotected queries, one looks for the threshold $b$ for which the query answer changes from 0 to 1. The attack is for IDP (Group IDP with $k=1$). 

It is important to note that, as discussed in Example \ref{ex:memberinference}, just because the $k$-laplace mechanism can leak more information about a query (such as $\tquery$) than if the query were not protected, this does not mean that our particular attack will benefit from it. Hence it is important to evaluate if there are inefficiencies in our attack.

\subsection{The Dataset}
As an illustration of the way the attack would be launched in practice, we use the well-known Banking dataset \cite{BankDataset} containing records of  45211 people. There are 7 numeric (integer) and 10 categorical attributes. As discussed in Section \ref{subsec:attackquery}, categorical attributes can be handled simply by encoding the values as integers (thus, for example, a yes/no attribute can be converted to an attribute whose values are 0 or 1).

Since part of the attack (Algorithm \ref{alg:columnreconstruct}) uses upper and lower bounds on the domain of numeric attributes, we choose the following conservative bounds:  $\left[-10^5, 10^6\right]$ for  \textbf{balance}, $\left[0, 10^4\right]$ for \textbf{duration}, $\left[-1, 2000\right]$ for \textbf{pdays} (-1 is a special coding for customers who were not previously contacted), $\left[0, 2000\right]$ for \textbf{previous}, $\left[0, 125\right]$ for \textbf{age}, $\left[0, 31\right]$ for \textbf{day}, and $\left[0, 100\right]$ for \textbf{campaign}. 

\subsection{Membership Inference}
In membership inference attacks, an attacker has uniquely identifying information about an individual and attempts to determine whether that individual is in the dataset (i.e., whether the number of people having the same values for those known attributes is 0 or 1). As explained in Example \ref{ex:memberinference}, when using the $k$-local Laplace mechanism to protect query answers, then no matter how small the privacy budget $\epsilon$ is, this attack succeeds with just two queries (no matter what the dataset is). If, on the other hand, queries are not protected at all, one simply asks whether the number of people with the known attributes is $> 0$. These results are summarized
in Table \ref{tab:member}.
\begin{table}[h!]
\begin{tabular}{|c|c|c|}\cline{2-3}
\multicolumn{1}{c|}{} &\textbf{Protected by IDP} & \textbf{No Protection}\\\hline
\textbf{\# Queries} &2 & 1\\\hline
\end{tabular}
\caption{Number of queries needed to launch a successful membership inference attack, no matter how small $\epsilon>0$ is, when queries are protected using the $k$-local Laplace mechanism of IDP vs. no protection at all.}
\label{tab:member}
\end{table}

\subsection{Attribute Inference}
We next consider an attacker who knows a person is in the data, has uniquely identifying information about the person, and is trying to discover additional attributes (like the person's \textbf{balance} in the banking dataset).

In this experiment, an attacker knows the following attributes about a target individual: \textbf{age}, \textbf{marital status}, \textbf{level of education}, \textbf{job type}, and whether the individual has a \textbf{housing loan}. These will be treated as the identifying attributes. The attacker could try to learn just one attribute (in this case it would be \textbf{balance}) or the attacker could try to learn the complete rest of the entire record. 

This attack can be carried out, for any arbitrarily low privacy loss budget $\epsilon>0$, as described in Example \ref{ex:att}. In the dataset, there were 1815 people who are unique on the linking attributes. In Table \ref{table:targeting} we show, on average, how many queries are needed to learn the balance attribute for those unique people, and how many queries are needed to learn the entire record.

    \begin{table}
   \begin{tabular}{|c|c|c|}\cline{2-3}
  \multicolumn{1}{c|}{} &\textbf{Protected by IDP} &   \textbf{No Protection}\\\hline
   ``Balance'' \# Queries & \textcolor{black}{29.9} & \textcolor{black}{29.9} \\\hline
  Full Record \# Queries & \textcolor{black}{131.2} & \textcolor{black}{131.2} \\\hline
  \end{tabular}
        \caption{Average number of queries to reconstruct the \textbf{balance} for people who are unique on linking attributes and the average number of queries to reconstruct all the non-linking attributes. Comparison between threshold range-count queries with and without IDP protection.}
        \label{table:targeting}
    \end{table}

To help interpret the numbers better, consider the \textbf{balance} attribute, for which we used lower and upper bounds of $-\text{\texteuro} 100,000$ and $\text{\texteuro}1,000,000$, which is a range that can be represented using $21$ bits. Thus, on average we need $\textcolor{black}{29.9/21\approx 1.4}$ queries per bit. This number can be further reduced if an attacker is not interested in the exact amount and only needs a few significant digits, or if the attacker already has a ballpark estimate of the target's balance.

Also note that \textbf{balance} was the attribute with the largest domain. The remaining 11 non-linking attributes are reconstructed using, on average, $\textcolor{black}{131.2-29.9=101.3}$ additional queries. We note that recovering a 
binary attribute $\att_{binary}$ is particularly straightforward. If we know someone is in the data and we know they are unique on a set of attributes $\att_1=a_1,\dots, \att_m=a_m$ then, if the value of $\att_{binary}=1$, there would only be one person in the data with $\att_1=a_1,\dots, \att_m=a_m, \att_{binary}=1$. Thus we can perform a membership inference attack with this combination of attributes and if the attack returns ``true,'' it means that $\att_{binary}=1$ for the target person, and if it returns ``false,'' then $\att_{binary}=0$. The cost of this is simply 2 ``protected'' queries.


\subsection{Dataset Reconstruction}
Efficient membership inference and partial/full record reconstruction for a target individual is already a strong demonstration of the exploitability of IDP. We next show that there are savings in bulk when performing full dataset reconstruction. That is, the \emph{average} number of queries \emph{per person} needed for reconstruction is less than the number of queries needed to attack a person individually because an attribute value may be shared by multiple people, so using one binary search to find this value and its count would produce results for multiple people at once. To take advantage of this type of bulk savings, we perform reconstruction starting with the binary attributes and then adding attributes to the reconstruction in order of the size of their domain.

We consider two types of experiments: how much effort is needed to reconstruct a single attribute, and how much effort is needed to reconstruct the entire dataset.

\subsubsection{Single Attribute Reconstruction}
Here we study how many queries are needed to reconstruct each attributes in isolation. That is, for each attribute, we are just interested in determining what are the distinct values that are present, and how many people have those values (in other words, we want to get an exact 1-way marginal). The number of queries depends on the number of unique values that appear for that attribute and are shown in Figure \ref{fig:onlyrow}.

\pgfplotsset{
  log x ticks with fixed point/.style={
  }
}

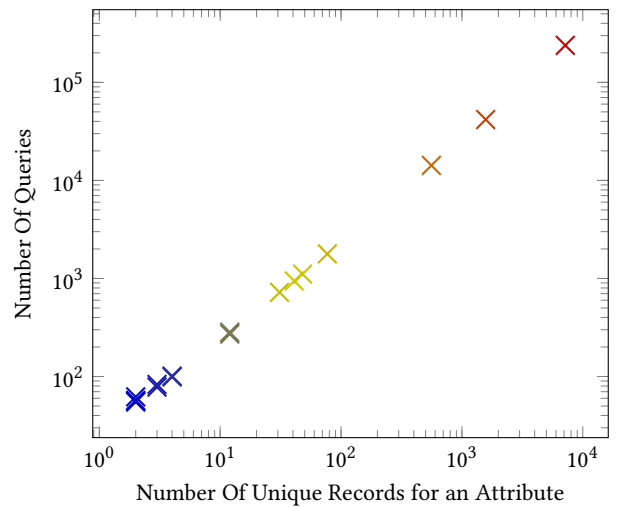
\begin{figure}[h!]
  \begin{center}
\begin{tikzpicture}
\begin{axis}[
    ymode=log,
    xmode=log,
    log x ticks with fixed point,
    xlabel=Number Of Unique Records for an Attribute,ylabel=Number Of Queries,
    legend style={at={(0.5,-0.2)},
	anchor=north,legend columns=-1},
]

\addplot[only marks,
    scatter, thick,
    mark=x,
    mark size=5pt]
table {
2   62  
2   57   
2   55  
2  56  
3   78  
3    83  
4   99  
4   101  
12   271  
12   283  
31  722  
41  943  
48  1109  
77  1781  
559  14227 
1573  41794 
7168  238462 
};  

\end{axis}
\end{tikzpicture}
  \caption{Number of queries needed to reconstruct an attribute in isolation. The plot indicates the number of queries vs. the number of distinct values that appear for the attribute.}\label{fig:onlyrow}
  \end{center}
\end{figure}

Note that reconstructing a binary attributes means determining exactly how many people had 0 (resp. 1) for that attribute and so requires a binary search that takes $O(\log(n))$ queries per attribute value (0 or 1), where $n$ is the number of people. The most difficult attribute to reconstruct as \textbf{balance}, which had 7168 distinct values in the dataset. Reconstruction required  \textcolor{black}{238,462} queries, which is approximately \textcolor{black}{33} queries per distinct value, or \textcolor{black}{5} queries per person.

\subsubsection{Entire Dataset Reconstruction}

The entire dataset consists of 17 attributes and contains 45,211 people, meaning that a full reconstruction is required to produce $17*45211=768587$ total items. Thus, one would expect that the number of queries would be in the millions. We allocated an $\epsilon=10^{-10}$ for each use of the $k$-local Laplace mechanism.
The results are shown in Table \ref{tab:fullrecon}.

    \begin{table}
   \begin{tabular}{|l|c|c|}\cline{2-3}
  \multicolumn{1}{c|}{} &\textbf{IDP Protection} &   \textbf{No Protection}\\\hline
   Total number of queries & \textcolor{black}{5,418,936} & \textcolor{black}{5,200,591} \\\hline
  Queries per person & \textcolor{black}{$\approx 119.9$} & \textcolor{black}{$\approx 115.0$} \\\hline
  Queries per data element & \textcolor{black}{$\approx 7.1$} & \textcolor{black}{$\approx 6.8$} \\\hline
  Total privacy budget spent & \textcolor{black}{$0.0005418936$} & N/A \\\hline
  \end{tabular}
  \caption{Full dataset reconstruction using threshold range-count queries, with and without the protection mechanisms of IDP. Each protected query access used $\epsilon=10^{-10}$ of the privacy budget.}\label{tab:fullrecon}
  \end{table}
  
Note that the total privacy budget spent (according to IDP privacy accounting) reconstructing the entire dataset was approximately $\textcolor{black}{0.0005}$. It can be made arbitrarily small. For example, if we used $10^{-11}$ per query, then the privacy budget would be 1/10th the size. In fact, for any target $\epsilon^*$, it is possible to guarantee that the total spent is at most $\epsilon^*$. For example, one could allocate $\epsilon_1=\min(\epsilon^*, 10^{-10})$ for the first query, $\epsilon_2=\epsilon_1/2$ for the next query, $\epsilon_3=\epsilon_2/2$ for the third query, and so on. This guarantees that the total spent is at most $\min(\epsilon^*, 10^{-10})$.

\section{Related Work}\label{sec:related}
\newcommand{\MYhref}[3][blue]{\href{#2}{\color{#1}{#3}}}%

Reconstruction attacks are possible when too many queries over confidential data are answered too accurately \cite{DN03}, or equivalently, when one tries to create a data product that supports all possible use-cases. This is not just a theoretical possibility, but also affects commercial offerings \cite{CN18}. 

Differential privacy, formally introduced in 2006 \cite{dmns06}, has been gaining steam as a mathematically rigorous privacy definition that protects against reconstruction and other embarrassing privacy attacks against public data products. This property allows organizations to use it to collect, protect and publish data products that would otherwise not be available at all.

There has been significant research on trying to improve the accuracy of the data products by carefully weakening the original differential privacy definition, while still preventing reconstruction. This includes approximate differential privacy \cite{ourdata}, concentrated differential privacy \cite{zcdp}, Renyi differential privacy \cite{renyidp}, and $f$-DP \cite{gaussdp}. These privacy definitions have  \emph{group privacy} guarantees which is what prevents reconstruction attacks \cite{Vadhan2017}.

There have, in fact, been numerous attempts to weaken differential privacy, strengthen it, and apply it to non-tabular data -- see the comprehensive comparative survey by Desfontaines and Pejó \cite{sokdps}.

One of the lines of research taken, which we call \emph{empirical neighbors} is spearheaded by IDP \cite{idp}. It is noteworthy for several reasons: (1) it was proposed by several long-term experts in privacy, (2) its flaws were not observed in the authoritative comparative survey \cite{sokdps} or the literature that cites IDP (e.g., \cite{prudence}), and (3) most notably, a group of prominent researchers, mostly from the economics field, called on the Census Bureau to stop using differential privacy and to explore alternatives \cite{hotzetal}. One approach, of adding noise depending on the local sensitivity (e.g., IDP) was deemed ``sensible'' as long as the local sensitivity itself is not explicitly revealed \cite[appendix c]{hotzetal}. In fact, their concern with local sensitivity was that it might not provide enough \emph{utility}.

It is known that adding noise based on local sensitivity does not satisfy differential privacy, and hence smooth sensitivity was proposed \cite{NRS07} and many believed that the main weakness of local sensitivity occurs when the local sensitivity is explicitly published \cite{hotzetal,Chetty_Friedman_2019}. However, the weaknesses we have demonstrated: queries that reveal more information when protected by IDP than if they had no protection at all (even when the local sensitivity is not explicitly published), and their use membership attacks, attribute inference, and full dataset reconstruction at arbitrarily low privacy costs (according to IDP privacy accounting) was not previously known, to the best of our knowledge.

Several authors investigated something similar to IDP, but as a diagnostic tool rather than a method for selecting mechanisms \cite{Charest_Hou_2017,redberg_per}. Here, a differentially private mechanism $\mech$ is chosen, it is applied to the dataset, and the privacy with respect to that dataset is studied after the fact. Charest et al. \cite{Charest_Hou_2017} used this methodology to compute an $\epsilon$ (this would be the $\epsilon$ that IDP would assign to $\mech$) and studied how well it correlates to the differential privacy $\epsilon$. They concluded that it was not a good estimate. Redberg and Wang \cite{redberg_per} studied how to make this ex-post analysis differentially private, so that the actual privacy cost of $\mech$ on the actual dataset could be revealed without breaching privacy.

\section{Conclusion}\label{sec:conc}
In this paper, we studied a class of privacy definitions called \emph{empirical} neighbors that condition on the observed data when choosing a mechanism. We showed that the preferred mechanisms can be exploited to reveal significant information about the true data. We also showed that the definitions themselves can be exploited to design mechanisms that directly leak private information. It is not clear whether this style of privacy definition can provide the right balance between privacy and utility in practice.

\begin{acks}
This work was supported by an NSF BAA award number 49100421C0022 and by NSF award CNS-1702760. 
\end{acks}


\bibliographystyle{ACM-Reference-Format}
\bibliography{refs}
\clearpage

\appendix
\section{Proofs from Section \ref{sec:idp}}
\printProofs[idp]

\section{Proofs from Section \ref{sec:bootstrap}}
\printProofs[bdp]

\section{Proofs from Section \ref{sec:generic}}
\printProofs[general]

\end{document}